\documentclass[JHEP,12pt]{article}
\usepackage{subcaption}
\usepackage{amssymb,amsmath}
\usepackage{xcolor}
\usepackage{braket}
\usepackage{jheppub}
\usepackage{endnotes}
\usepackage{comment} 
\usepackage[most]{tcolorbox}
\definecolor{block-gray}{gray}{0.85}
\newtcolorbox{oldversion}{colback=block-gray,grow to right by=0mm,grow to left by=0mm,boxrule=0pt,boxsep=0pt,breakable}

\renewcommand{\endnote}{\footnote}

\author[1]{Raphael Bousso,}
\author[2]{Xi Dong,}
\author[3]{Netta Engelhardt,}
\author[4]{Thomas Faulkner,} 
\author[5]{Thomas Hartman,}
\author[6]{Stephen  H. Shenker,}
\author[\: 6]{and Douglas Stanford}

\affiliation[1]{Center for Theoretical Physics and Department of Physics,\\
University of California, Berkeley, CA 94720, U.S.A.}
\affiliation[2]{Department of Physics, University of California, Santa Barbara, CA 93106, U.S.A.}
\affiliation[3]{Center for Theoretical Physics, Massachusetts Institute of Technology, \\Cambridge, MA 02139, USA}
\affiliation[4]{Department of Physics, University of Illinois Urbana-Champaign, Urbana, IL 61801, USA}
\affiliation[5]{Department of Physics, Cornell University, Ithaca, NY, USA}
\affiliation[6]{Stanford Institute for Theoretical Physics and Department of Physics,\\
Stanford University, Stanford, CA 94305, U.S.A. 
} 

\emailAdd{bousso@berkeley.edu}
\emailAdd{xidong@ucsb.edu}
\emailAdd{engeln@mit.edu}
\emailAdd{tomf@illinois.edu}
\emailAdd{hartman@cornell.edu}
\emailAdd{sshenker@stanford.edu}
\emailAdd{salguod@stanford.edu}

\begin{document}

\title{Snowmass White Paper: Quantum Aspects of Black Holes and the Emergence of Spacetime}

\abstract{

Black holes provide a window into the microscopic structure of spacetime in quantum gravity.  Recently the quantum information contained in Hawking radiation has been calculated, verifying a key aspect of the consistency of black hole evaporation with quantum mechanical unitarity.

This calculation relied crucially on recent progress in understanding the emergence of bulk spacetime from a boundary holographic description. Spacetime wormholes have played an important role in understanding the underpinnings of this result,
and the precision study of such wormholes, in this and other contexts, has been enabled by the development of low-dimensional models of holography.

In this white paper we review these developments and describe some of the deep open questions in this subject.  These include the nature of the black hole interior, potential applications to quantum cosmology, the gravitational explanation of the fine structure of black holes, and the development of further connections to quantum information and laboratory quantum simulation.
}

\maketitle

\section{Introduction}

 Nearly a half-century ago, Hawking showed that  black holes emit radiation~\cite{Hawking:1975vcx}, and ever since then the study of these objects has been a central part of our quest for a quantum theory of gravity.  Hawking's calculation also showed that the radiation left behind after black hole evaporation would be in a mixed (thermal) quantum state~\cite{Hawking:1976ra}, even if the initial state of the matter forming the black hole was pure. The fundamental irreversibility implied by this result sets up a sharp conflict between the unitary time evolution of quantum mechanics and general relativity at the event horizon. This ``black hole information paradox'' arises in a seemingly controlled regime characterized by a weakly curved geometry and low energy quanta, making it even more puzzling.

The AdS/CFT correspondence (also known as gauge/gravity duality)~\cite{Maldacena:1997re,Witten:1998qj,Gubser:1998bc} 
has provided us with a precise definition of nonperturbative quantum gravity, at least for a certain class of spacetimes, and therefore serves as a testing ground for the information paradox as well as other deep problems in quantum gravity. This holographic correspondence relates a non-gravitational quantum system, often a conformal field theory (CFT),  on the boundary of an asymptotically anti-de Sitter (AdS) spacetime 
 to a theory of quantum gravity (string or M theory) in the bulk of the spacetime.

 Many (but by no means all) of the entries in the dictionary relating bulk gravity and boundary quantum mechanics are known.  One basic entry links a high energy thermal state of the boundary system to a black hole in the bulk \cite{Witten:1998zw, Maldacena:2001kr}. 
 The entropy of the boundary system is the entropy of the black hole which, according to a result of Bekenstein and Hawking \cite{Bekenstein:1972tm,Hawking:1975vcx}, is determined by the area of its horizon. 
 
 Another entry that has played a central role in recent developments is the gravity dual of the entanglement entropy of a subregion of the boundary field theory.  Ryu and Takayanagi \cite{Ryu:2006bv,Hubeny:2007xt,Nishioka:2009un} argued that this is again proportional to the area of a surface, called the RT surface, ending on the boundary subregion, as illustrated in Figure \ref{fig:EW}. Both of these relations take the form
\begin{equation}\label{sintro}
 S = \frac{\mbox{Area}}{4G\hbar}  + \cdots
\end{equation}
for the entropy $S$, thus relating gravity, quantum mechanics, and information.  More generally, surface area constrains the amount of quantum information in a spacetime region, indicating that the holographic encoding of information in quantum gravity is not limited to the AdS/CFT correspondence.

The region enclosed by an RT surface --- or rather its generalization to include quantum effects, called a quantum extremal surface (QES)~\cite{Engelhardt:2014gca} --- is holographically encoded in the corresponding boundary subregion \cite{Bousso:2012sj,Czech:2012bh,Bousso:2012mh,Wall:2012uf,Headrick:2014cta,Dong:2016eik}. This encoding can be very complex, and is only partially understood. The problem of `decoding the hologram' is a research program known as bulk reconstruction that has illuminated deep links between quantum gravity and quantum information science. A striking example is the realization that information about the bulk is stored redundantly in the boundary theory via a quantum error-correcting code~\cite{Almheiri:2014lwa}.

Rapid progress on the information paradox has been made in the last three years, building on the Ryu-Takayanagi formula and many parallel developments in the theory of black hole information over the last decade. The initial breakthrough was made by locating a new, unexpected, quantum extremal surface in an evaporating black hole~\cite{Almheiri:2019psf,Penington:2019npb}. 
This enabled the first gravitational derivation of a key signature of unitarity --- the `Page curve' \cite{Page:1993wv} --- for the entropy of Hawking radiation.  The ideas underlying these calculations also extend the method of bulk reconstruction to black holes in more general spacetimes, taking a significant step beyond the AdS/CFT correspondence towards a theory of emergent spacetime in more realistic models of quantum gravity.

One important insight is that many of these results can also be understood from new nonperturbative saddle points in the semiclassical approximation to the gravitational path integral \cite{Almheiri:2019qdq,Penington:2019kki}.  Such semiclassical methods have a long history in this subject, going back to the evaluation of the entropy of a black hole from a Euclidean saddle point by Gibbons and Hawking \cite{Gibbons:1976ue}. 
In the Page curve calculation it turns out to be necessary to include the  contribution of spacetime wormhole geometries in the semiclassical analysis. 
 
An important catalyst for progress 
has been the construction of simple low dimensional models of quantum gravity where such wormhole effects can be studied in a controlled way.  These include the the Sachdev-Ye-Kitaev (SYK) model \cite{Sachdev:1992fk,KitaevTalks,Kitaev:2017awl,PhysRevB.59.5341,Maldacena:2016hyu,Sarosi:2017ykf} and its low energy limit, Jackiw-Teitelboim (JT) gravity \cite{Jackiw:1984je,Teitelboim:1983ux} in two dimensions.

The above threads share two common traits: a dramatic acceleration of progress over the past decade, and increasingly deep and central connections to quantum information science and quantum many-body physics.  This progress is gratifying but many mysteries remain.   What is the bulk dual of a typical state in the boundary system, and how is this related to the firewall paradox~\cite{Almheiri:2012rt} that helped initiate these developments?  What is the nature of the black hole singularity and what role does it play in this circle of ideas?  How do these ideas extend beyond AdS spacetimes, especially to cosmologies resembling our world?  What is the bulk explanation for the individual microstates of a black hole?  Is it possible to construct model systems in the lab that would actually allow us to gain experimental insight into some of these issues?

\section{Recent progress}

\subsection{Emergence of spacetime}\label{subsec:emergst}

\begin{figure}
\centering
\includegraphics[width=0.3\textwidth]{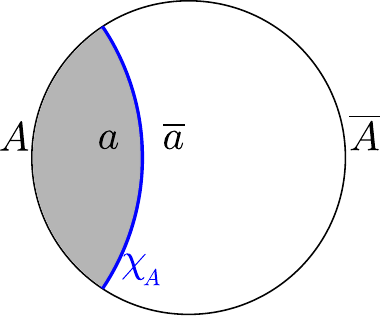}
\caption{Subregion-subregion duality relates a boundary subregion $A$ with its entanglement wedge $a$ (of which only a time slice is shown here).  The entanglement wedge is bounded by a (quantum) extremal surface $\chi_A$.}
\label{fig:EW}
\end{figure}

Several related dualities\footnote{The chronology includes matrix models and 2d gravity \cite{David:1984tx,Ambjorn:1985az,Kazakov:1985ds,Kazakov:1985ea} in the double-scaled limit \cite{Douglas:1989ve,Brezin:1990rb,Gross:1989vs}, M(atrix) theory \cite{Banks:1996vh}, and AdS/CFT \cite{Maldacena:1997re,Gubser:1998bc,Witten:1998qj}. See \cite{Itzhaki:1998dd,Polchinski:1999br} and \cite{McGreevy:2003kb} for connections.} have connected quantum gravity systems to non-gravitational degrees of freedom. In each case, the gravitational spacetime emerges from the collective behavior of the nongravitational degrees of freedom. 

This idea is sharpest in the context of AdS/CFT, but even there, the basic mechanism has only been clarified recently -- insights from a quantum information perspective have been central to these developments.  In particular a dictionary has been developed relating bulk gravitational quantities to quantum information-theoretic quantities.

These ideas grew out of the  Ryu-Takayanagi \cite{Ryu:2006bv} proposal connecting the areas of minimal (RT) surfaces to entanglement entropies of boundary subregions.   This proposal has been extensively developed \cite{Hubeny:2007xt,Nishioka:2009un,Headrick:2010zt,Dong:2013qoa,Camps:2013zua,Dong:2016fnf,Dong:2016hjy,Freedman:2016zud,Takayanagi:2017knl,Chen:2018rgz,Faulkner:2018faa,Dutta:2019gen}.  In particular, quantum corrections have been understood and a semiclassical gravitational path integral derivation has been formulated \cite{Lewkowycz:2013nqa,Faulkner:2013yia,Barrella:2013wja,Faulkner:2013ana,Engelhardt:2014gca,Jafferis:2015del,Dong:2017xht}.  A key notion is the entanglement wedge of a boundary subregion \cite{Bousso:2012sj,Czech:2012bh,Bousso:2012mh,Wall:2012uf,Headrick:2014cta}.  This is a spacetime region in the bulk that is bounded by a (quantum) extremal surface \cite{Engelhardt:2014gca} (classically, an RT surface) associated with the boundary subregion, as shown in Figure \ref{fig:EW}.

A large body of work has led to the central concept of subregion-subregion duality \cite{Bousso:2012sj,Czech:2012bh,Bousso:2012mh,Wall:2012uf,Headrick:2014cta,Dong:2016eik}: the quantum information present in a subregion of the boundary field theory is exactly the information needed to describe the bulk quantum state in the entanglement wedge.  In particular, bulk operators in the entanglement wedge of a boundary subregion can be reconstructed as some boundary operators on that subregion \cite{Dong:2016eik}.
  
\begin{figure}
\centering
     \hspace*{\fill}
     \begin{subfigure}[b]{0.2\textwidth}
         \centering
         \includegraphics[width=\textwidth]{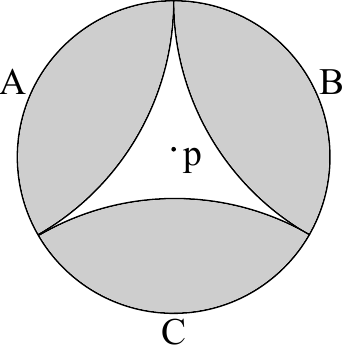}
     \end{subfigure}
     \hfill
     \begin{subfigure}[b]{0.2\textwidth}
         \centering
         \includegraphics[width=\textwidth]{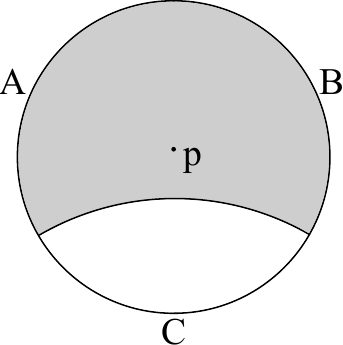}
     \end{subfigure}
     \hfill
     \begin{subfigure}[b]{0.2\textwidth}
         \centering
         \includegraphics[width=\textwidth]{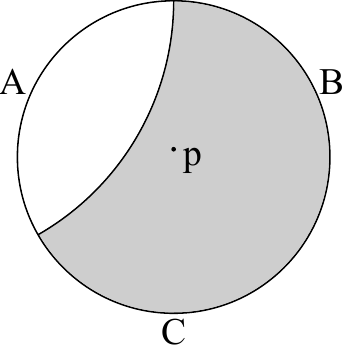}
     \end{subfigure}
     \hfill
     \begin{subfigure}[b]{0.2\textwidth}
         \centering
         \includegraphics[width=\textwidth]{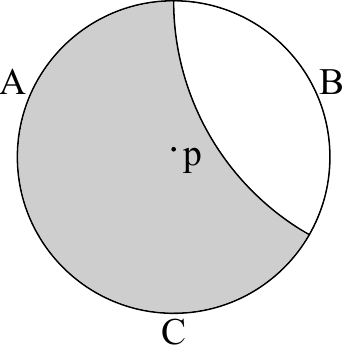}
     \end{subfigure}
     \hspace*{\fill}
\caption{The ``ABC'' puzzle for reconstructing a bulk operator at point $p$ in three different ways is resolved by the insight that holography works as a quantum error-correcting code.  The first panel shows the individual entanglement wedges of the three regions $A,B,C$.  None of these contains the point $p$ so it cannot be reconstructed from information in a single region.   The second, third and fourth panels illustrate the entanglement wedges of the regions $A \cup B$, $B \cup C$, and $A \cup C$ respectively. Each one of these regions contains $p$ so they provide the data for  distinct, redundant, reconstructions of $p$.}

\label{fig:abc}
\end{figure}

  A striking aspect of subregion-subregion duality  is that the information about the bulk is stored redundantly in the boundary.  It functions as a quantum error-correcting code \cite{Almheiri:2014lwa}. 
A simple example of this is the ``ABC'' puzzle illustrated in Figure \ref{fig:abc}: a bulk operator at point $p$ can be reconstructed using three different boundary operators\endnote{This follows from the AdS-Rindler version of the Hamilton-Kabat-Lifschytz-Lowe (HKLL) \cite{Hamilton:2006az} procedure of reconstructing bulk operators in terms of nonlocal boundary operators by solving the bulk equations of motion as operator equations.} supported on the distinct, different regions $A \cup B$, $B \cup C$, and $A \cup C$.  This redundant quantum encoding is the basic mechanism at work in quantum error-correcting codes.\endnote{Similar error-correcting properties have been realized in concrete toy models of holography built from tensor networks \cite{Swingle:2009bg,Pastawski:2015qua,Hayden:2016cfa,Donnelly:2016qqt,Qi:2018shh}. Also see \cite{Mintun:2015qda,Freivogel:2016zsb,Almheiri:2016blp,Harlow:2016vwg,Pastawski:2016qrs,Cotler:2017erl,Hayden:2018khn,Almheiri:2018xdw,Akers:2018fow,Dong:2018seb,Yoshida:2018ybz,Bao:2018pvs,Chen:2019gbt,Faist:2019ahr,Akers:2019wxj,Akers:2019nfi,Dong:2019piw,Kim:2020cds,Faulkner:2020iou,Akers:2020pmf,Faulkner:2020hzi,Faulkner:2020kit} for further work on quantum error correction and information recovery in holography.}

Another approach to bulk reconstruction that highlights the  emergence of bulk locality and causality involves the concept of modular flow \cite{Haag:1992hx,Casini:2008cr,Witten:2018zxz}.  This is a generalized notion of time evolution that treats the logarithm of a general density matrix as a Hamiltonian, also known as the modular Hamiltonian.  This generalizes the Rindler time evolution near the black hole horizon generated by the ordinary CFT Hamiltonian \cite{Bisognano:1976za,Hislop:1981uh,Casini:2011kv}.
A result that is basic to this program is that the bulk and boundary modular Hamiltonians can be identified up to an area term \cite{Jafferis:2015del}.
 
Operator reconstruction with modular flow allows one to reach operators everywhere inside the entanglement wedge \cite{Faulkner:2017vdd}, thus in many situations reaching behind causal horizons.\endnote{The Petz map \cite{petz2007quantum} is a specific reconstruction map used in quantum error-correcting codes and also involves a kind of modular flow, achieving a similar outcome \cite{Almheiri:2014lwa,Cotler:2017erl,Chen:2019gbt,Penington:2019kki} and thus connecting these approaches to bulk reconstruction.}

A deeper connection between modular flow and bulk emergence follows from studying relations between bulk causality and analyticity in modular time \cite{Faulkner:2018faa,DeBoer:2019kdj}.\endnote{See \cite{Lin:2019qwu,Leutheusser:2021qhd} for related approaches to bulk reconstruction.}
For example, by studying analyticity of correlation functions in modular time 
one can establish a connection between the emergence of local bulk physics and the saturation of \emph{modular chaos} bounds constraining the growth of these correlators. Such bounds are connected to the chaos bound \cite{Maldacena:2015waa}, which constrains the growth of finite temperature out-of-time-ordered correlators. It is well known that gravity saturates the chaos bound, and that this is connected to the existence of a smooth black hole horizon. The modular chaos generalization attempts to move these results away from black hole horizons to general locations in spacetime. 

Constraints from analyticity in modular time are also known to directly constrain dynamical aspects of the emergent gravitational physics, most notably via an interesting connection to quantum energy conditions \cite{Bousso:2015wca,Balakrishnan:2017bjg}. These conditions generalize the well known null energy condition that plays an important role in constraining the causal structure of classical general relativity and underlies the celebrated Penrose singularity theorem \cite{Penrose:1964wq}.  

Causal aspects of semi-classical gravity, which includes leading order quantum corrections,  are often usefully constrained by the Quantum Focusing Conjecture \cite{Bousso:2015mna,Bousso:2015wca,Koeller:2015qmn,Leichenauer:2018obf,Balakrishnan:2019gxl}.
For example, the Quantum Focusing Conjecture can be used to prove a basic causal constraint on entanglement wedge reconstruction - that bulk regions should nest when the corresponding boundary regions nest \cite{Wall:2012uf,Akers:2016ugt}.
As shown in \cite{Faulkner:2018faa}, entanglement wedge nesting
is indeed connected to analyticity in modular time, a further constraint on the correlation functions discussed above that arises for theories saturating the modular chaos bound. 
This line of reasoning has been particularly fruitful, leading to proofs of energy conditions in QFT \cite{Balakrishnan:2017bjg,Ceyhan:2018zfg} without gravity.\footnote{Further applications of modular flow to these issues can be found in \cite{Chen:2018rgz,Engelhardt:2018kcs,Dutta:2019gen,Bousso:2019dxk,Bousso:2020yxi,Levine:2020upy}.}

These developments have highlighted a deep connection between AdS/CFT and a formal mathematical framework called Algebraic Quantum Field Theory, whose constructs are very natural from the quantum information theoretic point of view.  This abstract approach has provided powerful new tools and new insights into these issues.\endnote{See for example \cite{Casini:2008cr,Witten:2018zxz,Longo:2018zib,Casini:2019kex,Faulkner:2020hzi,Furuya:2020tzv,Witten:2021unn}. }

\subsection{The information paradox}\label{subsec:infopara}
\begin{figure}
\begin{center}
\includegraphics[width=8.6cm]{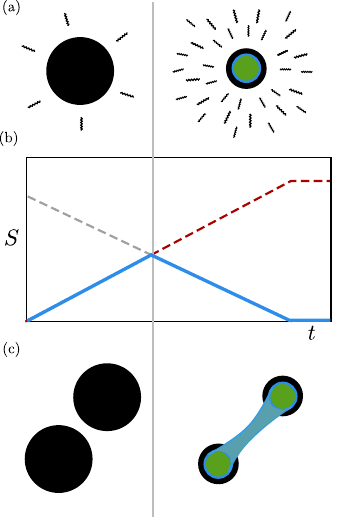}
\end{center}
\caption{Progress on the information paradox. (a) Black hole evaporation. An island (green) bounded by a quantum extremal surface (blue) appears at late stages in the evaporation. (b) The von Neumann entropy of Hawking radiation: Hawking's calculation (dashed red) leads to a paradox when the radiation entropy exceeds the black hole entropy (dashed gray). If black hole evaporation is unitary, then the true entropy should follow the Page curve (blue). (c) Two copies of the black hole can be used to probe the quantum information in the radiation. At late times, spacetime wormholes join the black hole interiors. This leads to the creation of islands, which in turn produce the unitary Page curve. \label{fig:pagecurve}}
\end{figure}

In defiance of a longstanding expectation, new calculations~\cite{Penington:2019npb, Almheiri:2019psf} have produced striking evidence that low energy, semiclassical gravity can detect unitarity in black hole evaporation (see \cite{Almheiri:2020cfm} for a review). 
These recent developments are summarized in Figure \ref{fig:pagecurve}. A sharp diagnostic of the information paradox is the von Neumann entropy of Hawking radiation,
\begin{equation}
S_{\rm R} = - \mbox{tr}\, \rho_{\rm R} \ln \rho_{\rm R} \ ,
\end{equation}  
where $\rho_{\rm R}$ is the density matrix of the radiation. 
Hawking's calculation indicates that $\rho_{\rm R}$ is mixed and that  $S_{\rm R}$ grows monotonically as the black hole evaporates, while the Bekenstein-Hawking entropy of the black hole,
\begin{equation}
S_{\rm BH} = \mbox{Area}/4G \ ,
\end{equation}
decreases to zero. This leads to a contradiction: Unitarity requires $S_{\rm R} \to 0$ at late times, because if the initial state of the black hole is pure then the final state of the radiation must be pure also. Furthermore, general properties of entangled quantum systems require $S_{\rm R} \leq S_{\rm BH}$. Since Hawking's calculation takes place in a region of low curvature, this contradiction constitutes an apparent violation of effective field theory in a regime where strong quantum gravity corrections should be suppressed.

The behavior of the radiation entropy expected from unitary evaporation is known as the Page curve \cite{Page:1993wv} (Figure \ref{fig:pagecurve}b). It is this universal curve that was determined quantitatively from recent semiclassical calculations using the quantum extremal surface (QES) formula for entropy in quantum gravity~\cite{Engelhardt:2014gca}:
\begin{equation}\label{eq:QES}
    S_{R}=\frac{\mathrm{Area}[\chi]}{4 G} +S_{\mathrm{bulk}}\equiv S_{\mathrm{gen}}[\chi]
\end{equation}
Here $\chi$ is a QES: a surface where $S_{\rm gen}$ is stationary under small perturbations. (If there are multiple stationary surfaces, $\chi$ is taken to be the one with minimal entropy.) The second term $S_{\mathrm{bulk}}$ is the von Neumann entropy of quantum fields in a region bounded by $\chi$.\endnote{The region must also be homologous to the region $R$ containing the radiation, and $S_{\mathrm{bulk}}$ is by definition the von Neumann entropy calculated \textit{without} gravitational instantons. From an information-theoretic viewpoint it is not the von Neumann entropy of the exact density matrix of the quantum fields, but of a density matrix defined on the code subspace appropriate to bulk reconstruction.} In many situations the quantum correction $S_{\mathrm{bulk}}$ has a small effect compared to the area term (which determines the classical RT surface). But at sufficiently late stages of the evaporation, when according to the Page curve the entropy $S_R$ should begin to decrease, $S_{\mathrm{bulk}}$ has a dramatic effect. A new QES in the black hole interior becomes dominant, and its effect is to produce exactly the decreasing part of the Page curve. 

The QES formula therefore agrees with unitary black hole evaporation, but in fact, it provides much more than just a formula for the entropy. An extension of the bulk reconstruction ideas discussed in Section \ref{subsec:emergst} leads to the conclusion that the region behind the QES --- an `island' in the black hole interior --- is actually encoded in the Hawking radiation~\cite{Penington:2019npb,Almheiri:2019hni}.\endnote{This is a concrete realization of earlier intuitions like $\mathrm{A} = \mathrm{R_B}$ and $\mathrm{ER = EPR}$ \cite{Bousso:2012as,Nomura:2012sw,Verlinde:2012cy,Papadodimas:2012aq,Maldacena:2013xja} relating the radiation to the interior.  It provides a geometric  realization of the Hayden-Preskill protocol \cite{Hayden:2007cs}, whereby the information in the black hole interior can be decoded from the radiation at late times. Concrete reconstruction procedures have been discussed using the Petz map~\cite{Penington:2019kki} and modular flow~\cite{Chen:2019iro}.
}

The island can be traced back to spacetime wormholes~\cite{Coleman:1988cy,Giddings:1988cx,Maldacena:2004rf} in the gravitational path integral \cite{Penington:2019kki, Almheiri:2019qdq}. This can be described in the context of the von Neumann entropy, but it is simpler to use the quantum purity, $\mbox{tr}\, \rho_R^2$. This is a convenient diagnostic for unitarity because in a pure state, $\mbox{tr} \rho_R^2 = 1$, while in a mixed state, $\mbox{tr} \,\rho_R^2 < 1$. Because it involves two copies of the density matrix, the purity is calculated by a gravitational path integral involving two copies of the black hole known as replicas (Figure \ref{fig:pagecurve}c).
At early times, the purity agrees with Hawking's calculation of monotonically increasing entropy. However, at late times there is a dynamical transition to another saddle point in the gravitational path integral in which the two black holes are joined through their interiors by a spacetime wormhole. In the wormhole phase the purity is larger, and indeed it returns to unity as the black hole evaporates. This is consistent with a unitary final state. In the analogous replica calculation of the von Neumann entropy, the mouth of the wormhole becomes the island inside the black hole, and an evaluation of the wormhole action justifies the QES formula.\endnote{This idea was initially tested in a doubly-holographic model~\cite{Almheiri:2019qdq} in which the evaporating black hole was holographically dual to a higher-dimensional purely classical bulk. The QES formula without islands was derived earlier from the gravitational path integral in \cite{Lewkowycz:2013nqa,Barrella:2013wja,Faulkner:2013ana,Dong:2016hjy,Dong:2017xht}.}

The appearance of the island in the entanglement wedge indicates that at late stages of the evaporation the black hole interior is encoded in the Hawking radiation.\footnote{For an alternative perspective, and a discussion of subtleties that arise in separating the radiation from the black hole, see \cite{Laddha:2020kvp,Geng:2020fxl,Geng:2021hlu,Raju:2021lwh,Geng:2021mic}.} However, the relation between the island and the radiation constitutes a departure from the standard holographic dictionary: data in the island is not spatially connected to the radiation, so reconstruction is necessarily more subtle.
A natural question, then, is what qualitatively sets apart the island -- and more generally the deep black hole interior -- from the rest of the bulk in terms of reconstruction. This question is at the root of decoding the black hole interior from the radiation.   

Recent developments have pointed to the central role of quantum computational complexity in understanding this issue.
General considerations \cite{Hayden:2007cs,Harlow:2013tf,Aaronson:2016vto,Kim:2020cds} imply that decoding the Hawking radiation is exponentially complex. More precisely, the state of the Hawking radiation cannot be reliably distinguished from a random state by any  quantum circuit whose size is polynomial in the black hole entropy. 

Ideas about tensor networks~\cite{Swingle:2009bg,Hartman:2013qma} and the geometrization  of quantum complexity~\cite{Susskind:2014rva,Susskind:2014jwa, Stanford:2014jda,Roberts:2014isa,Brown:2015lvg,Bouland:2019pvu, Susskind:2020gnl}
have led to a conjecture geometrizing reconstruction complexity in terms of QESs.  The so-called Python's Lunch conjecture states that reconstruction of bulk data in the entanglement wedge is exponentially complex if that data lies behind a subdominant (non-minimal) QES~\cite{Brown:2019rox}.\footnote{The converse is also expected to be true~\cite{Engelhardt:2021mue, Engelhardt:2021qjs}, which has implications for a holographic understanding of Hawking's calculation.} This perspective  unifies the developing geometric picture of QESs and both dominant and subdominant saddles of the gravitational path integral with earlier developments on the process and feasibility of decoding the Hawking radiation. 

These calculations have opened new avenues of research in black hole information that are reshaping our approach to the information paradox. At the same time, they have raised many new questions about the interplay of quantum mechanics and gravity, some of which are discussed in Section \ref{sec:futuredir}.

\subsection{Wormholes, low-dimensional gravity, and the SYK model}\label{subsec:lowdim}
Some of the phenomena discussed above depend crucially on subtle nonperturbative effects in the gravitational path integral, for example  the replica wormhole saddles that compute the Page curve.  Such effects are difficult to study in a controlled way in general.   A strategy employed extensively in recent years has been to use simple two-dimensional gravity models to study these effects, as well as other aspects of the quantum physics of black holes, with precision.

An important step here was the development of the Sachdev-Ye-Kitaev (SYK) model \cite{Sachdev:1992fk,KitaevTalks,Kitaev:2017awl,PhysRevB.59.5341,Maldacena:2016hyu}, an ensemble of simple but strongly-interacting quantum mechanical systems.  Concretely, the SYK model describes the quantum mechanics of a collection of $N$ Majorana fermions coupled with generic four-fermion couplings which are drawn from a probabilistic ensemble.  Among other things these systems display the maximally chaotic, fast scrambling behavior characteristic of black holes \cite{Hayden:2007cs,Sekino:2008he,Shenker:2013pqa,Shenker:2013yza,KitaevTalkBreakthrough,Shenker:2014cwa,Maldacena:2015waa}.  

At low energies these systems are described by a two-dimensional gravity theory (Jackiw-Teitelboim or ``JT'' gravity \cite{Jackiw:1984je,Teitelboim:1983ux,Almheiri:2014cka,Jensen:2016pah,Maldacena:2016upp,Engelsoy:2016xyb}) that provides a universal description of near-extremal black holes. In the SYK model, this theory arises by a concrete change of variables starting from the quantum mechanics of Majorana fermions. This precise mapping has made it possible to sharpen the connections between gravity and quantum mechanics, to resolve some puzzles, and to generate new ones. In gravity variables, several of the areas of progress have involved the physics of wormholes.

One example of this is the role of  replica wormholes in addressing the black hole information paradox.  Another concerns the statistics of black hole energy levels, packaged together into a convenient function called the spectral form factor \cite{Haake:1315494,Papadodimas:2015xma}\hspace{.15em}\endnote{The spectral form factor is related to the two-point correlation function with the operator matrix elements removed. For work on wormholes and correlation functions see \cite{Blommaert:2019hjr,Saad:2019pqd}.} 
\begin{equation}\label{SFF}
Z(\beta+i t)Z(\beta-it) = \sum_{n,m} e^{-\beta(E_n + E_m)}e^{it(E_n - E_m)}~.
\end{equation}
From the perspective of black hole physics, the LHS is computed by a pair of black hole geometries, and the answer apparently decreases forever as a function of $t$. But in a true quantum system, the RHS cannot decay forever; instead, it will rattle around erratically as the phases oscillate. This conflict is known as Maldacena's black hole information problem \cite{Maldacena:2001kr}.  Numerical studies of the SYK model and the concrete mapping to gravity showed that the lack of decay of the RHS arises due to a spacetime wormhole that can connect the two black holes together \cite{Saad:2018bqo}.  The functional form of this late time behavior is a signature of the random matrix statistics of the energy levels, which are a universal property of quantum chaotic systems \cite{Haake:1315494} and so should be a feature of  black holes more broadly.\endnote{More precisely, the wormhole describes the ramp part of the spectral form factor.   The ultimate late time behavior, the plateau, has a more complicated origin.} 

More precisely, this wormhole accurately computes the answer after averaging over the ensemble of SYK theories, smearing the erratic oscillations into a smooth function. This suggests a connection between simple gravity theories and ensembles of quantum systems, as exemplified by the exact duality between dilaton gravity theories like JT and quantum theories where the Hamiltonian is drawn from a  random matrix ensemble.\endnote{This duality is formally analogous to the older random matrix description of low-dimensional string theories, reviewed in \cite{Ginsparg:1993is,DiFrancesco:1993cyw,Seiberg:2004at}.   But here the perspective is different -- the random matrix is the full boundary system Hamiltonian, not a single field in a field theory.} So wormholes solved one puzzle but created another: how can gravity describe a single quantum system, rather than an ensemble?\endnote{In fact the connection between wormholes and ensembles is an old one, going back to work on baby universes and wormholes in the 1980s \cite{Coleman:1988cy,Giddings:1988cx}.   For a modern reformulation of these ideas in the AdS/CFT context see \cite{Marolf:2020xie}.}
We will discuss this issue further in Section \ref{subsec:noisy}.

Another application of wormholes involves the physics of entanglement. Black holes manifest certain patterns of entanglement via geometrical connections of spatial wormholes.\endnote{The basic example of this is the Einstein-Rosen bridge in the eternal Schwarzschild black hole.  This is the bulk embodiment of entanglement in of the boundary thermofield double state \cite{Maldacena:2001kr}.} Entanglement alone does not allow signalling, which is dual to the statement that the corresponding wormholes are not traversable. But a small interaction between the two black holes can lead to traversability \cite{Gao:2016bin}. This phenomenon was studied in the SYK model and in JT gravity \cite{Maldacena:2017axo}, and eventually used to construct theoretical examples of traversable wormholes in our own four-dimensional world  \cite{Maldacena:2018lmt,Maldacena:2018gjk,Maldacena:2020sxe,Fu:2019vco}. It has also created new connections to quantum information theory and quantum simulation. From a quantum information perspective, passing through the wormhole is a particular implementation of quantum teleportation, carried out by an elegant protocol invented by gravity itself. This protocol has been used to inspire and explain experiments carried out using noisy quantum simulators \cite{Brown:2019hmk,Schuster:2021uvg,Blok:2020may}.

Progress has not been limited to the physics of wormholes, however. For example, precise computations of the density of states in JT gravity \cite{Cotler:2016fpe, Bagrets:2017pwq, Stanford:2017thb,Mertens:2017mtv,Kitaev:2018wpr, Yang:2018gdb} were used in  \cite{Iliesiu:2020qvm,Heydeman:2020hhw} to resolve an old problem \cite{Preskill:1991tb} related to the energy spectrum of near-extremal black holes.

\section{Future directions}\label{sec:futuredir}

\subsection{Behind the horizon}
Classically the black hole horizon serves as a sharp division of spacetime into  regions that are accessible and inaccessible to a distant observer.  Deep questions exist about the nature of the region behind the horizon -- the black hole interior.

\subsubsection{The firewall paradox}

 In 2007, Hayden and Preskill \cite{Hayden:2007cs} used modern tools from quantum information theory to show that information that falls into an old black hole should rapidly become recoverable from the radiation. In 2013, Almheiri, Marolf, Polchinski, and Sully (AMPS) \cite{Almheiri:2012rt} used this result to strengthen a previous argument of Mathur \cite{Mathur:2009hf}, deriving a paradox that motivated them to conjecture that the geometry of an old black hole should be dramatically altered behind the horizon: a ``firewall'' would form.\endnote{These ideas were partially anticipated in \cite{Braunstein:2009my}.} Whether this actually happens is still an open question.\endnote{There has been extensive work on trying to resolve this question.  Examples of approaches include \cite{Bousso:2012as,Nomura:2012sw,Verlinde:2012cy,Papadodimas:2012aq,Maldacena:2013xja,Mathur:2013gua}.}  However the island results discussed above  suggest that gravity may evade the original entanglement-based argument by encoding the black hole interior in the radiation \cite{Bousso:2012as,Nomura:2012sw,Verlinde:2012cy,Papadodimas:2012aq,Maldacena:2013xja}.

Other arguments \cite{Almheiri:2013hfa,Marolf:2013dba} for firewalls attempt to establish their existence in random states in  Hilbert space, and these arguments are not obviously affected by such an identification. So an apparently sharp question remains completely open: what is the interior of a black hole in a random quantum state? Is the interior  even uniquely determined by the state?

\subsubsection{The black hole singularity}
Putting firewalls aside, it is still certainly the case that classical geometry breaks down behind the horizon near the black hole singularity~\cite{Penrose:1964wq,Wall:2010jtc}.
Ever since the discovery of the Schwarzschild solution the nature of this singularity has been a mystery.   The developments described in this white paper have not, to date, cast new light on this problem and it remains a central task of a theory of quantum gravity to understand it. 

Classically the singularity represents an ``end of time" behind the horizon, raising deep questions about the nature of ordinary quantum time evolution there.\endnote{Time evolution outside the horizon proceeds without end and joins to the manifestly unitary quantum time evolution of the boundary theory.}  
The intriguing black hole final state proposal~\cite{Horowitz:2003he,Gottesman:2003up,StanfordTalk,AlmheiriTalk}
posits that the quantum state of the black hole  is postselected to match a fixed final state at the singularity.   This allows for a unitary black hole scattering matrix, but represents a deviation from ordinary quantum mechanics in the interior.  This highlights a key question:  what is the proper quantum description of bulk dynamics inside the horizon?

\subsection{Cosmology}

Up till now this white paper has focused on black holes and AdS spacetimes,   but the geometry of our universe is quite different.  Its cosmology at both early and late times seems to be an exponentially expanding one,  consistent with de Sitter, not Anti-de Sitter, space.

Finding a nonperturbative description of quantum gravity in such cosmologies is a problem of central importance -- some current approaches to this problem will be discussed in a separate white paper on cosmology and string theory.
Here we will content ourselves with pointing out a few  areas where the ideas discussed in this paper may be of some relevance to this problem.

A basic reason to expect a connection to black holes is the existence of horizons in de Sitter space.  An observer sees a horizon whose area defines a de Sitter entropy and which semiclassical calculations show to radiate thermally \cite{Gibbons:1977mu} at the de Sitter temperature \cite{Figari:1975km}. These thermal fluctuations are related to the primordial fluctuations that are directly observed via the cosmic microwave background (CMB).

One crucial difference between de Sitter space and black holes is that  de Sitter observers are in the interior of their horizons.   A second difference is the absence of an analog of the black hole singularity that observers encounter in the future.  (There is a past singularity though, the big bang.)
Despite these differences we can still ask whether QES notions continue to be useful here. Do islands play a role? How about wormholes?

There are cosmologies with future singularities  (``big crunches") that observers eventually encounter.  These cosmologies typically have spherical regions where the matter entropy is large compared to the classical area \cite{Fischler:1998st,Bousso:1999xy}, suggesting that matter entanglement could perhaps compete with classical geometry~\cite{Bousso:2021sji}, as in the island effect in black hole evaporation.  And there are models where islands, replica wormholes and other related configurations,  bra-ket wormholes, do appear \cite{Maldacena:2019cbz,Cotler:2019nbi,Anous:2020lka,Dong:2020uxp,Krishnan:2020fer,Chen:2020tes,Hartman:2020khs,VanRaamsdonk:2020tlr,Aguilar-Gutierrez:2021bns,Shaghoulian:2021cef,Teresi:2021qff,Shaghoulian:2022fop}. 
It is an open question whether they occur in the more realistic, expanding cosmologies.  

There are more basic questions.  What degrees of freedom does the de Sitter entropy count and what does this concept mean in a geometry with exponentially expanding space? Bulk reconstruction, as it is presently understood, relies on a boundary region or auxiliary degrees of freedom that exist outside of the gravitating spacetime and which, through their entanglement structure, encode the properties of the emergent geometry. In a closed universe, or in an expanding cosmology with no spacelike boundary, there is no clear separation between gravitating and non-gravitating regions, so this crutch must be modified or abandoned.   There are a number of proposals for formulating a holographic description, but as of now they are incomplete. Further exploration of holography in this setting is an important task for the future.

\subsection{Fine structure of black holes}\label{subsec:noisy}

The boundary quantum description of a finite entropy black hole has  a discrete spectrum of energy levels
-- each one of these levels describes a microstate of the black hole.   A complete description of the bulk must include a description of these states, but the way this is realized in general is a mystery.\endnote{Such a bulk description has been found for the microstates of certain extremal supersymmetric black holes, as well as a few nonextremal examples \cite{Lunin:2001jy,Lin:2004nb,Bena:2015bea,Jejjala:2005yu,Ganchev:2021pgs}.  The active ``fuzzball" program \cite{Mathur:2005zp,Mathur:2009hf,Skenderis:2008qn,Guo:2021blh} seeks to build on this success.    The bulk realization of a typical state of a large non-extremal black hole is currently unknown though, and there are reasons to suspect that the general case will be qualitatively different from the extremal one.}
 It seems likely that understanding this description will require substantial new insights, and will shed important new light on the nature of quantum gravity.

This discreteness causes quantum noise \cite{1997PhRvL..78.2280P,1999JPhA...32.6903H,Barbon:2014rma,Cotler:2016fpe,Stanford:2020wkf} in addition to the signal computed using gravitational structures, like wormholes.  In certain quantities, like the Page curve, we expect that this noise will be very small. But in others, like the ramp in the spectral form factor or individual matrix elements of the radiation density matrix $\rho_R$, or of the black hole  S matrix itself,  we expect the noise to be comparable to the signal.  The simplest way to isolate the gravitational contribution is by averaging over an ensemble of boundary quantum systems. And in fact 2D JT gravity is precisely dual to such an ensemble, in this case a random matrix ensemble \cite{Saad:2019lba,Stanford:2019vob}.

Such ensembles of theories can appear to violate basic rules of quantum mechanics~\cite{Bousso:2019ykv,Bousso:2020kmy}; for example, observables in completely disjoint universes can nonetheless be correlated by the ensemble average. In gravitational theories this correlation is captured by wormholes.
This sharp tension between  decoupled boundary theories and bulk geometries that connect them is referred to as the factorization problem \cite{Witten:1999xp,Maldacena:2004rf}. Its resolution is likely to be related to an understanding of black hole microstructure.

These ideas raise several important questions.
First, we do not expect the usual examples of holographic duality to be precisely described by an ensemble -- the boundary theories are too special.\endnote{An argument for the absence of a bulk ensemble has been presented in \cite{McNamara:2020uza}.}$^,$\endnote{There are other ways to average over the noise that apply to systems with a fixed boundary Hamiltonian.   Averaging over time intervals on the ramp of the spectral form factor is an example. Such averaged quantities are the ones that are expected to display universal random matrix behavior in quantum chaotic systems.}   What ingredients need to be added to the bulk gravitational description to describe the noise?  A number of proposals have been discussed \cite{Blommaert:2019wfy,Marolf:2020xie,Eberhardt:2020bgq,Eberhardt:2021jvj,Saad:2021rcu,Saad:2021uzi,Blommaert:2021fob,Almheiri:2021jwq,Mukhametzhanov:2021nea}, but the full story remains to be told.

 Second, how much does the semiclassical gravitational description itself know about this noise? To what extent do wormholes in e.g.,~four-dimensional gravity provide a useful statistical description of it \cite{Cotler:2020ugk,Mahajan:2021maz,Belin:2020hea}? Are there further examples of precise dualities between ensembles of boundary systems and bulk gravitational ones \cite{Afkhami-Jeddi:2020ezh,Maloney:2020nni,Collier:2021rsn,Benjamin:2021wzr,Meruliya:2021utr,Benjamin:2021ygh,Heckman:2021vzx}? What lessons can be drawn from thinking about these questions from the perspective of the Hilbert space of baby universes \cite{Coleman:1988cy,Giddings:1988cx, Polchinski:1994zs,Marolf:2020xie,Marolf:2020rpm,Marolf:2021ghr}?

\subsection{New connections to quantum information }
\subsubsection{The holographic code}

Holography packages the quantum state of the bulk theory into the boundary theory according to a type of quantum error correcting code. Such codes must have remarkable properties, with the flexibility and precision to describe all of bulk physics. By trying to explain how basic features of the bulk theory are represented, we can try to reverse-engineer these codes and extract lessons both for quantum gravity and the theory of quantum error correction.

Some progress has been made on this problem recently. For example, non-flat entanglement spectra can be explained using codes with central commutative algebras \cite{Harlow:2016vwg,Akers:2018fow,Dong:2018seb,Dong:2019piw}. Internal symmetries have been incorporated \cite{Harlow:2018tng,Faist:2019ahr}, and we are beginning to understand how the entanglement wedge and its area can emerge from properties of the code \cite{Harlow:2016vwg,Akers:2021fut}. State specific entanglement wedges have been understood \cite{Hayden:2018khn,Akers:2020pmf}.

However, many puzzles remain, ranging from structural questions like the origin of local physics on sub-AdS scales \cite{Hayden:2016cfa} and the compatibility of bulk and boundary dynamics, to more detailed questions like the apparent ability of holography to defeat location-based cryptography with polynomial  entanglement resources 
\cite{May:2019yxi,May:2019odp}.

\subsubsection{String theory and quantum information}

While ideas from quantum information have been given natural dual descriptions in semi-classical gravity, understanding the role of string theory in this duality is an important open question. As a well-established UV completion of gravity one might have expected it to play a larger role. For small string lengths the RT area formula gets corrections from higher derivative couplings in the effective action which  have been extensively studied. However these perturbative corrections are not always sufficient to capture all stringy effects even at small string lengths. Indeed, just as stringy effects can be enhanced near black hole horizons due to boosted kinematics, they can also be enhanced when reconstructing operators inside the entanglement wedge \cite{Chandrasekaran:2021tkb}.

In holography, string theory with nonzero string length offers an interpolation between emergent spacetime (for small string length) and more weakly-coupled boundary degrees of freedom (at large string length). How does the relationship between quantum information and spacetime play out along this axis? For example, can we study entanglement entropy and bulk reconstruction in string theory? Entanglement requires a notion of splitting a system -- how do we locally split the string theory Hilbert space?  To what extent can we understand the RT formula as arising from stringy edge modes?\endnote{Examples of work on this problem include \cite{Dabholkar:1994ai,He:2014gva,Witten:2018xfj,Hartnoll:2015fca,Donnelly:2016jet,Balasubramanian:2018axm,Hubeny:2019bje}.}  This connects to old ideas \cite{Susskind:1994sm} about explaining black hole entropy as strings ending on the horizon.  A related idea is the possible transition between the large number of highly excited string states and black hole states, as a parameter is varied \cite{Bowick:1985af,Susskind:1993ws,Sen:1995in,Horowitz:1996nw,Horowitz:1997jc,Chen:2021dsw}. Understanding this transition might shed light on the nature of black hole microstates and the various factorization puzzles that arise from using semiclassical gravity to describe the dual of entangled states of decoupled systems~\cite{Marolf:2012xe,Harlow:2015lma,Guica:2015zpf, Harlow:2018tqv}.

\subsection{Quantum gravity in the lab}
Quantum gravity has traditionally been a largely theoretical field.
However, increasing control over laboratory quantum systems and near-term (NISQ) quantum  devices may make it possible to simulate quantum systems with interesting holographic duals: ``quantum gravity in the lab.'' Progress in this direction has started already with \cite{Brown:2019hmk,Schuster:2021uvg,Blok:2020may}. Following this example, we can expect that ideas from quantum gravity might help to inspire or explain experiments. An equally exciting prospect is that experiments on simulated quantum systems might challenge or inform our understanding of theoretically intractable (strongly coupled) limits of quantum gravity systems, with possible relevance to black hole physics, high energy scattering, or the early universe.

\section{Outlook}
Over the last decade progress in the areas discussed here has been impressively rapid.  Cross-fertilization from different fields like quantum information theory and many-body physics has contributed new vitality to the subject, and has helped cause new avenues of research
to multiply.   As is often the case in rapidly developing interdisciplinary subjects,  progress has come from surprising, unexpected directions.    
The continuing ferment in this field makes us optimistic that the coming decade will be similarly productive and surprising.

\section*{Acknowledgements}
RB is supported in part by the Berkeley Center for Theoretical Physics; by the Department of Energy, Office of Science, Office of High Energy Physics under QuantISED Award DE-SC0019380 and under contract DE-AC02-05CH11231; and by the National Science Foundation under Award Number 1820912.
XD is supported in part by the Air Force Office of Scientific Research under award number FA9550-19-1-0360 and by funds from the University of California.
NE is supported in part by NSF grant no. PHY-2011905, by the U.S. Department of Energy Early Career Award DE-SC0021886, by the U.S. Department of Energy grant DE-SC0020360 (Contract 578218), by the John Templeton Foundation and the Gordon and Betty Moore Foundation via the Black Hole Initiative, and by funds from the MIT department of physics. TF is supported by the Air Force Office of Scientific Research under award number FA9550-19-1-0360 and by the Department of Energy under award number DE-SC0019183.
TH is supported by the Simons Foundation and NSF grant PHY-2014071.   SS is supported in part by NSF grant PHY-1720397. DS is supported in part by DOE grant DE-SC0021085 and by the Sloan Foundation.

\bibliographystyle{JHEP}
\bibliography{bibliography}

\end{document}